# Unusual Exchange Bias Associated with Phase Separation in Perovskite Cobaltites

Young Sun, Yan-kun Tang, and Zhao-hua Cheng,

*State Key Laboratory of Magnetism, Institute of Physics, Chinese Academy of Sciences,*
*Beijing 100080, P. R. China*

We report the observation of unusual exchange bias phenomena in the doped perovskite cobaltites $La_{1-x}Sr_xCoO_3$ (x = 0.15, 0.18, and 0.30) in which a spontaneous phase separation into ferromagnetic clusters embedded in a non-ferromagnetic matrix occurs. When the $La_{1-x}Sr_xCoO_3$ samples are cooled in a static magnetic field through a freezing temperature, the magnetization hysteresis loops exhibit exchange bias, *i.e.*, the loops shift to the negative field and the magnetization becomes asymmetric. Moreover, exchange bias disappears when the measuring magnetic field is high enough. These results suggest that the intrinsic phase inhomogeniety in a spontaneously phase-separated system may induce an interfacial exchange anisotropy after a field cooling. The diminution of exchange bias in high magnetic fields is ascribed to the propagation of ferromagnetic regions with increasing magnetic field.

The exchange bias phenomenon refers to a shift of the magnetization hysteresis loop away from zero field due to a unidirectional anisotropy. This anisotropy is usually caused by the exchange coupling at the interface between ferromagnetic (FM) and antiferromagnetic (AFM) spin structures after the system is cooled in a static magnetic field through the Néel temperature of the AFM [1,2]. So far the research on exchange bias has been mainly focused on FM/AFM thin films where a well defined and controllable interface exists [3-5]. In addition to FM/AFM interfaces, exchange bias has also been observed in other types of interfaces involving a ferrimagnet (FI) (e.g., FI/AFM, FI/FM) [6,7] or involving a spin glass (SG) phase (e.g., FM/SG, AFM/SG, and FI/SG) [8-11]. These interfaces are usually achieved through artificially designed phase inhomogeneity, for example, by making artificial magnetic bilayers or mixing magnetic particles with a different matrix (granular systems). In this letter, we report the first observation of exchange bias phenomenon in spontaneously phase separated perovskite oxides in which intrinsic phase inhomogeneity plays a crucial role.

Hope-doped perovskite oxides, such as manganites and cobaltites, have drawn a lot of research attention since the early of 1990s, mainly due to the discovery of the colossal magnetoresistance (CMR) effect in them. Recent progress in CMR materials has reached the conclusion that intrinsic phase separation should play a crucial role in understanding their peculiar physical properties [12,13]. Especially, recent studies have shown that the hope-doped cobaltites such as $La_{1-x}Sr_xCoO_3$ exhibit a particularly clear form of phase separation for a broad range of doping level x, as evidenced by many experimental results obtained using various techniques including electron microscopy [14,15], nuclear magnetic resonance (NMR) [16-18], and small-angle neutron scattering (SANS) [19]. It has been well recognized that the phase separation in $La_{1-x}Sr_xCoO_3$ is in the form of coexistence of FM metallic regions and non-FM insulating regions. Furthermore, with this form of phase separation, $La_{1-x}Sr_xCoO_3$ perovskites exhibit hysteretic magnetoresistance with temperature and field dependence characteristic of inter-granular giant magnetoresistance (GMR) [19]. Therefore, it has been proposed that the spontaneously phase separated $La_{1-x}Sr_xCoO_3$ system is analog to the artificial granular films that are composed of FM particles embedded in a non-FM matrix [19].

Although the scenario of phase separation in $La_{1-x}Sr_xCoO_3$ has been well established, little attention has been put on the interfaces between the intrinsic inhomogeneous phases. In this work, we report the observation of unusual exchange-bias phenomenon associated with phase separation in $La_{1-x}Sr_xCoO_3$. This finding suggests that in a spontaneously phase-separated system the exchange coupling at the interfaces between the FM regions and the surrounding non-FM phase may create a unidirectional anisotropy (exchange anisotropy) when the sample is cooled in a static magnetic field. Moreover, unlike conventional exchange bias phenomena, exchange bias associated with phase separation can be completely removed in high magnetic fields due to the growth of FM regions with increasing magnetic field.

Polycrystalline $La_{1-x}Sr_xCoO_3$ (x = 0.15, 0.18, and 0.30) samples were prepared with solid state reaction method. A stoichiometric mixture of $SrCO_3$, $Co_3O_4$, and $La_2O_3$ powders was well ground and calcined twice at 800 and 950 $^o$C for 24 h. Then, the resulting powder was pressed into pellets and sintered at 1100 and 1150 $^o$C for 24 h, respectively. X-ray diffraction patterns show that all samples are single phase with rhombohedral structure. The magnetization measurements were performed using a commercial Quantum Design SQUID magnetometer.

Previous studies have established the phase diagram of $La_{1-x}Sr_xCoO_3$. Intrinsic phase separation,

*i.e.*, coexistence of FM and non-FM phases, occurs within a broad doping level (0.1< x < 0.5) [16]. For x < 0.18, the non-FM phase dominates and the small FM regions are well isolated so that the system is insulating. At the critical composition (x = 0.18), a metal-insulator transition occurs probably due to the percolation of isolated FM metallic regions. For x > 0.18, the FM phase becomes dominating and the resistivity exhibits a metallic behavior. In this work, we focused on the critical composition (x = 0.18) where the competition between FM and non-FM phase is the most significant, with complemental data of x = 0.15 and 0.30 samples.

Fig. 1 shows the temperature dependence of magnetization in a low magnetic field ($H$ = 10 Oe) with the zero-field cooled (ZFC) and field cooled (FC) processes for $La_{0.82}Sr_{0.18}CoO_3$. At 240 K the magnetization starts to increases rapidly with decreasing temperature, implying the onset of FM ordering. However, there is not a well-defined Curie temperature, $T_c$, because of the broadness of the rise. This indicates that the system is not in a long range FM ordering state but in the phase separated state in which the FM clusters and non-FM regions coexist. Below 240 K, the ZFC and FC magnetization separate. As illustrated in the inset of Fig. 1, the ZFC magnetization exhibits a peak around 140 K, which could be due to the freezing of the moments of FM clusters in random directions (similar to the blocking effect in magnetic particles). In addition, there could be another contribution to the freezing of moment. It has been well known that a spin-disordered interface/surface layer is usually formed when a FM particle is embedded in a non-FM matrix [10] or the magnetic particle size is small enough (the finite size effect) [9]. Therefore, it would be expected that spin-disordered glassy regions could exist at the interfaces between the FM clusters and the non-FM matrix. In fact, Co NMR has revealed that glassy regions coexist with FM and non-FM regions in $La_{1-x}Sr_xCoO_3$ [16,17]. These interfacial glassy regions may contribute to the freezing of moments as well as other glassy behaviors of $La_{1-x}Sr_xCoO_3$.

As we mentioned above, an interface involving a FM and a spin glass structure may cause exchange bias phenomenon. Therefore, it would be interesting to explore if the intrinsic interfaces associated with phase separation could give rise to exchange bias. To clarify this point, we have measured the hysteresis loops of $La_{0.82}Sr_{0.18}CoO_3$ at 5 K with both the ZFC and FC processes. For the ZFC process, the sample was cooled in zero magnetic field from room temperature to 5 K. For the FC process, the sample was cooled in 10 kOe magnetic field from room temperature to 5 K. Then the hysteresis loops were measured between ± 10 kOe. As shown in Fig. 2, while the ZFC magnetization has a normal hysteresis loop centered at zero field, it is clear that the FC hysteresis loop shifts to the negative field and the magnetization becomes asymmetric. The magnitude of the shift is known as exchange bias, $H_E = (H_{c1} + H_{c2})/2$, where $H_{c1}$ and $H_{c2}$ are the left and right coercive field, respectively. With the coercive fields obtained from the FC loop, we get the value of exchange bias, $H_E \approx -500$ Oe. The FC hysteresis loop also has an increased coercivity, $H_C = |H_{c1} - H_{c2}|/2 = 2395$ Oe, compared to the coercivity of 2310 Oe for the ZFC loop. All these behaviors are the characteristics of exchange bias phenomenon.

The appearance of exchange bias in $La_{0.82}Sr_{0.18}CoO_3$ indicates that a unidirectional anisotropy is built up after the sample is cooled in a magnetic field. We argue that this anisotropy is due to interfacial exchange interaction associated with phase separation. As we already pointed out above, the phase-separated $La_{1-x}Sr_xCoO_3$ is analog to the artificial granular films that are composed of FM particles embedded in a non-FM matrix. For those artificial granular systems, exchange bias has been widely observed [11,20], and it can be qualitatively understood with the following picture. When the sample is cooled in the presence of a magnetic field, an energy favorable spin configuration at the interface will be selected through the exchange interaction between the spin glass phase and the FM particles/clusters. The exchange interaction competes with thermal fluctuation. As a result, below the freezing temperature, the interfacial spin configuration is frozen and becomes more stable at lower temperatures. When the measuring field is reversed, the spins of FM particles start to rotate. However, the spin configuration at the interface may remain unchanged. Therefore, the spins at interface exert a microscopic torque on the FM spins to keep them in their original position. Thus, the field needed to reverse the FM spins will be larger because an extra field is required to overcome the microscopic torque. However, when the field is rotated back to its original direction, the FM spins will start to rotate at a smaller field due to the interfacial interactions which now exert a torque in the same direction as the field. Based on this picture, exchange bias in $La_{0.82}Sr_{0.18}CoO_3$ can be qualitatively interpreted in terms of the interfacial coupling associated with intrinsic phase inhomegeniety. We also measured the FC hysteresis loops at different temperatures. As shown in Fig. 3, exchange bias of $La_{0.82}Sr_{0.18}CoO_3$ decays with increasing temperature and eventually disappears above the freezing temperature, ~ 140 K, consistent with above picture.

In order to further understand exchange bias associated with phase separation, we also measured the hysteresis loops with different measuring magnetic fields. For each measurement, the sample was cooled in a field of 3 kOe from room temperature to 5 K, then the hysteresis loops were measured between ± 3, ± 5, ± 10, ± 30, and ± 50 kOe, respectively. The results are shown in Fig. 4. When the measuring field is low (H ≤ 10 kOe), the FC hysteresis loops always shift to the negative field and the magnetization becomes asymmetric with descending and ascending field, suggesting that a unidirectional anisotropy exists after the field cooling. However, when the measuring field is high enough, H ≥ 30 kOe, the FC hysteresis loops become symmetric and do not show any shift, *i.e.*, exchange bias disappears in high magnetic fields. In any case, the ZFC hysteresis loops do not show shift whatever the measuring magnetic field is.

This peculiar feature of exchange bias in $La_{0.82}Sr_{0.18}CoO_3$ distinguishes a phase-separated system from an artificial granular system. Unlike an artificial granular system where the size of FM particles is independent of applied magnetic field, the size of FM regions in phase-separated $La_{1-x}Sr_xCoO_3$ is strongly dependent on applied magnetic field. With increasing applied magnetic field, the FM clusters grow up [14]. Therefore, the interfaces between the FM regions and the surrounding non-FM matrix are moving as the FM clusters expand. With the increment of the size of FM clusters, the relative proportion of the interface layers to the FM clusters significantly decreases. Once the FM cluster is big enough, the small portion of interface spins can not pining the huge moments of the FM region. Consequently, no exchange anisotropy exists. This effect is similar to the well known $1/t_{FM}$ law of exchange bias in FM/AFM bilayers [1], *i.e.*, exchange bias is inversely proportional to the thickness, $t_{FM}$, of the FM layer, and becomes negligible when the FM layer is too thick. Meanwhile, as the reversed measuring field is strong enough, the spin configuration at the interfaces set up under the positive cooling field could be destroyed completely, which can also lead to the disappearance of exchange bias. In a word, the spontaneous interfaces associated with phase separation are not so stable in high magnetic field due to the propagation of FM phase. A high magnetic field can disturb the interfacial exchange coupling and remove the exchange anisotropy.

Since spontaneous phase separation occurs in $La_{1-x}Sr_xCoO_3$ for a broad doping level (0.1 < x < 0.5), it would be expected that exchange bias may appear in other $La_{1-x}Sr_xCoO_3$ samples.

Especially, at low x level the proportion of FM phase is low, and consequently the FM clusters are small and well isolated. In this situation, the interfacial exchange coupling is expected to have a more significant effect. As the average size of FM clusters is smaller and the propagation of clusters could be much slower, exchange bias in $La_{1-x}Sr_xCoO_3$ with $x < 0.18$ should remain till a higher magnetic field than that in $La_{0.82}Sr_{0.18}CoO_3$.

To examine this view, we have studied the hysteresis loops of two other samples, $La_{0.85}Sr_{0.15}CoO_3$ and $La_{0.70}Sr_{0.30}CoO_3$. In Fig. 5, we show the FC hysteresis loops at 5 K for $x = 0.15$ and 0.30 samples with a cooling field of 3 kOe. Both samples exhibit exchange bias. Moreover, as we have expected, for $La_{0.85}Sr_{0.15}CoO_3$ where the FM clusters are small and well isolated, exchange bias remains even in the measuring field of 50 kOe. In contrast, for $La_{0.70}Sr_{0.30}CoO_3$ where the FM phase dominates although phase separation still exists, exchange bias is only observable when the measuring magnetic field is less than 15 kOe. Thus, the above picture of exchange bias associated with phase separation has been further confirmed.

In summary, we have discovered unusual exchange bias phenomena associated with phase separation in $La_{1-x}Sr_xCoO_3$. When the samples are cooled in a magnetic field through a freezing temperature, the magnetic hysteresis loops shift to the negative field and the magnetization becomes asymmetric. Moreover, exchange bias decays with the increment of temperature and measuring magnetic field. These results suggest that, in a spontaneously phase-separated system, the exchange coupling at the interfaces between the FM regions and the surrounding non-FM phase may create an exchange anisotropy when the sample is cooled in a static magnetic field. Peculiarly, unlike conventional exchange bias phenomenon, exchange bias associated with phase separation can be completely removed in a high magnetic field, depending on the doping level, due to the propagation of FM phase with increasing magnetic field.

The authors are grateful to Prof. Jian-wang Cai for helpful discussion and Prof. Tong-yun Zhao for assistance in experiments. This work was supported by the State Key Project of Fundamental Research and the National Natural Science Foundation of China.

Figure captions

Fig. 1 Temperature dependence of magnetization with ZFC and FC processes for $La_{0.82}Sr_{0.18}CoO_3$. The inset illustrates a freezing temperature of 140 K with ZFC process for $La_{0.82}Sr_{0.18}CoO_3$.

Fig. 2 Hysteresis loops of $La_{0.82}Sr_{0.18}CoO_3$ at 5 K measured after zero-field cooling and field cooling in 10 kOe field. $H_{c1}$ and $H_{c2}$ refer to the left and right coercive field, respectively. Exchange bias and asymmetric magnetization appear when the sample is cooled in a magnetic field.

Fig. 3 The FC hysteresis loops of $La_{0.82}Sr_{0.18}CoO_3$ at different temperatures. Exchange bias decays with increasing temperature and eventually disappears above the freezing temperature, ~ 140 K.

Fig. 4 The FC (in 3 kOe) hysteresis loops at 5 K with different measuring magnetic fields. Exchange bias shows up for low measuring fields, but it disappears when the measuring field exceeds 30 kOe.

Fig. 5 The FC (in 3 kOe) hysteresis loops of $La_{0.85}Sr_{0.15}CoO_3$ and $La_{0.70}Sr_{0.30}CoO_3$ at 5 K. Both exhibit exchange bias. While it remains even in 50 kOe for $La_{0.85}Sr_{0.15}CoO_3$, exchange bias is observable only when the measuring field is below 15 kOe for $La_{0.7}Sr_{0.3}CoO_3$.

Fig.1

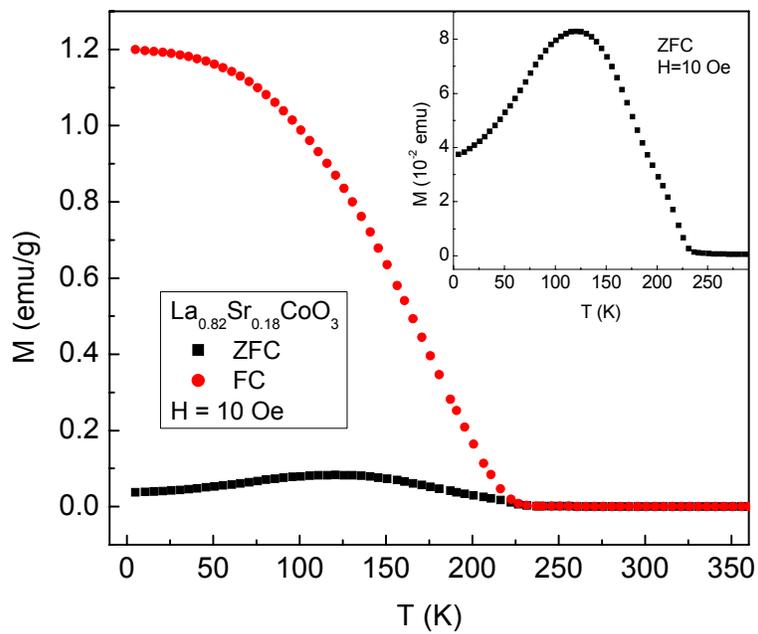

Fig. 2

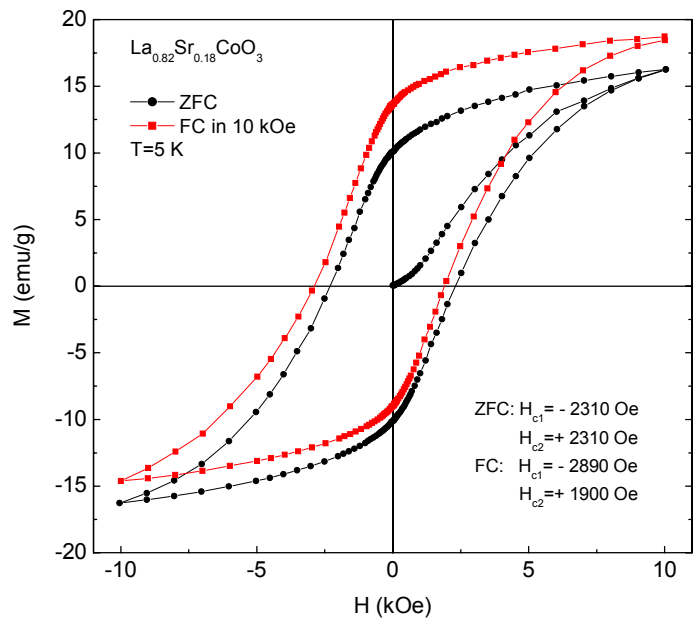

Fig. 3

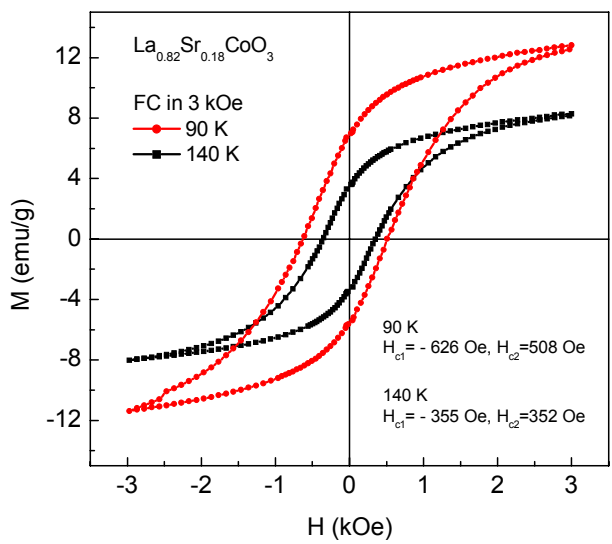

Fig. 4

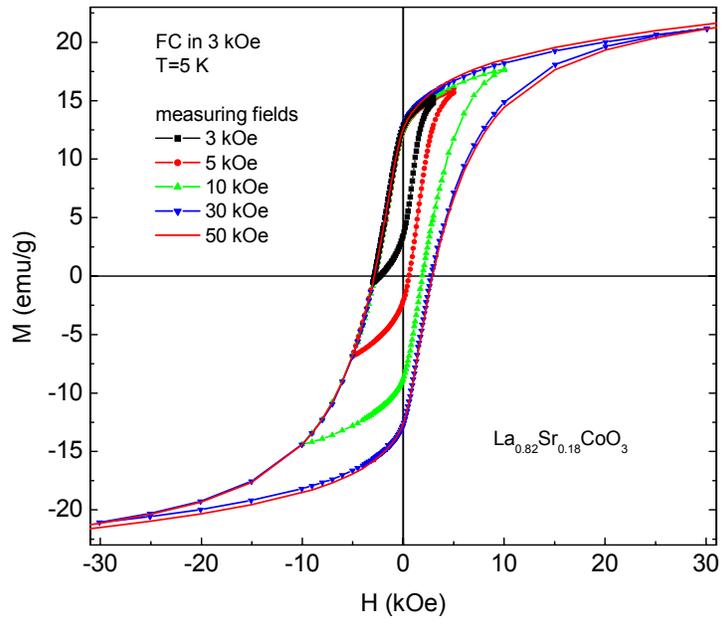

Fig. 5

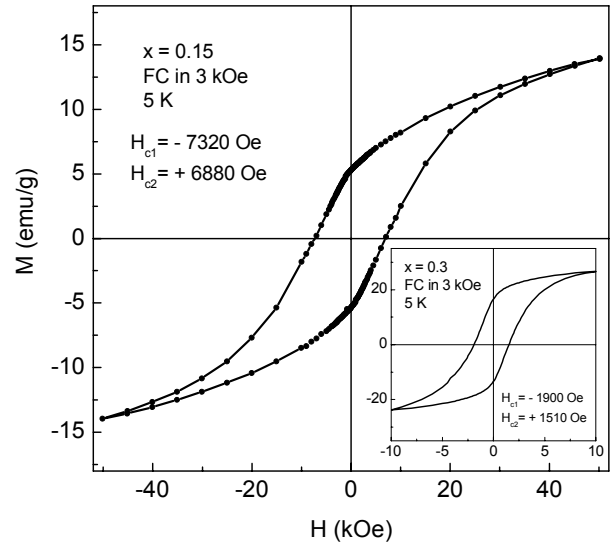